\documentclass[preprintnumbers,amsmath,amssymbm,prd]{revtex4}
\usepackage{epsfig}
\usepackage{graphicx}
\usepackage{amssymb}

\begin{document}
\title{Cloudy black-hole configurations of the composed Einstein-Yang-Mills-Higgs field theory: Analytic treatment}
\author{Shahar Hod}
\affiliation{The Ruppin Academic Center, Emeq Hefer 40250, Israel}
\affiliation{ }
\affiliation{The Jerusalem Multidisciplinary Institute, Jerusalem 91010, Israel}
\date{\today}

\begin{abstract}
\ \ \ Non-Abelian black holes that violate the non-hair conjecture 
are known to exist in the Einstein-Yang-Mills-Higgs field theory. 
The onset of hair formation in this composed physical system is marked by the presence of `cloudy' 
black-hole spacetimes, magnetically charged Reissner-Nordstr\"om black holes that support spatially 
regular linearized field configurations. 
In the present paper we explore, using {\it analytical} techniques, 
the physical and mathematical properties of the composed black-hole-linearized-field 
bound-state configurations of the Einstein-Yang-Mills-Higgs theory. 
In particular, we derive a remarkably compact analytical formula for
the discrete resonance spectrum $\{\mu_k\}^{k=\infty}_{k=1}$ of the dimensionless mass-radius 
parameter $\mu\equiv m_{\text{w}} r_+$ that characterizes the cloudy 
black-hole-field configurations in the regime $1\ll r_+\ll m^{-1}_{\text{w}}$ 
of small-mass weakly-magnetized black holes (here $r_+$ is the horizon radius of the central supporting black hole 
and $m_{\text{w}}$ is the mass of the Yang-Mills field due to the Higgs effect). 
Interestingly, our results reveal that the excited black-hole-field resonances are characterized by 
the asymptotic $k$-independent dimensionless ratio ${{\mu_{k+1}}/{\mu_{k}}}=e^{-{{2\pi}/{\sqrt{3}}}}$. 
\end{abstract}
\bigskip
\maketitle

\section{Introduction}

The no-hair conjecture \cite{Whee,Car} asserted that black holes are unable to support 
non-trivial, spatially regular static matter configurations outside their absorbing horizons. 
Motivated by this physically influential idea, a series of mathematically rigorous no-hair theorems 
has been established in the physics literature (see \cite{Chas,Hart,BekVec,BekMay,Her1n,Hodstationary} 
and references therein). 
These mathematically elegant theorems provide definitive proofs that static black-hole spacetimes with spatially 
regular horizons and asymptotically flat boundary conditions cannot support minimally coupled scalar, spinor, and 
massive vector fields.

Interestingly, subsequent studies of the non-linearly coupled Einstein-matter field 
equations demonstrated that the no-hair conjecture does not hold universally. 
In particular, it has been explicitly proved in the physically important works \cite{fir1,fir2,fir3} that hairy black-hole spacetimes exist 
within the framework of the composed Einstein-Yang-Mills field theory 
(for a comprehensive review, see \cite{VolGal}). 
It was later shown (see \cite{hairSky,hairHig2,hairstr,Hod1,Hod2,Her1,Her2,Gar1,Hersc1,Hersc2,Hodsc1,aaa1,aaa2,Moh,GV,GV2,Hodlm} 
and references therein) that, contrary to the original no-hair conjecture, several Einstein–matter theories admit black-hole configurations endowed with external hairy fields. 

The composed Einstein-Yang-Mills-Higgs system has attracted considerable attention during the last three 
decades (see \cite{EYMH1,EYMH2,EYMH3,VolGal} and references therein). 
In particular, using numerical techniques it has been revealed \cite{EYMH1,EYMH2,EYMH3,VolGal} 
that the Einstein-Yang-Mills-Higgs theory is characterized by the existence of 
magnetically charged black holes that support nontrivial external hairy configurations. 
The onset of hair formation 
in this composed physical system is signaled by the existence of `cloudy' 
black-hole spacetimes \cite{Notelc,Hod1,Hod2,Her1,Her2,Gar1}, 
magnetically charged Reissner-Nordstr\"om black holes with spatially 
regular horizons that support linearized bound-state matter configurations of the Yang-Mills field.

The physically interesting numerical results presented in \cite{EYMH1,EYMH2,EYMH3,VolGal} indicate 
that the dimensionless mass-radius parameter $m_{\text{w}} r_+$ that characterizes the 
cloudy black-hole-field configurations of the composed Einstein-Yang-Mills-Higgs theory 
has a discrete 
spectrum $\{[m_{\text{w}} r_+]_k\}^{k=\infty}_{k=0}$ of resonant eigenvalues (Here 
$\{r_+,m_{\text{w}}\}$ are respectively the 
horizon radius of the central black hole and 
the mass of the spatially regular supported Yang-Mills field \cite{Notema}). 

The main goal of the present paper is to explore the physical and mathematical properties of
the cloudy back-hole-field configurations that mark 
the critical boundary between bald black-hole spacetimes and composed black-hole-field hairy configurations 
in the Einstein-Yang-Mills-Higgs theory. 
Remarkably, we shall explicitly prove below that, in the dimensionless 
regime \cite{Noteunits,NoteP1,Noteannu}
\begin{equation}\label{Eq1}
m_{\text{w}} r_+\ll1\ \ \ \ \text{with}\ \ \ \ r_+\gg1\  ,
\end{equation}  
the cloudy black-hole-field bound-state 
configurations of the composed Einstein-Yang-Mills-Higgs system are amenable to 
an {\it analytical} treatment.  

In particular, in the present paper we shall derive, using analytical techniques,  a 
compact formula for the excited resonance spectrum 
$\{[m_{\text{w}} r_+]_k\}^{k=\infty}_{k=1}$ that characterizes the cloudy black-hole-field bound-state 
configurations of the  
Einstein-Yang-Mills-Higgs theory in the regime (\ref{Eq1}) of small-mass weakly-magnetized 
black holes \cite{Notewa1,Notewa2}. 

\section{Description of the system}

We shall analyze the physical and mathematical properties of the bound-state 
Reissner-Nordstr\"om-black-hole-linearized-Yang-Mills-field cloudy 
configurations that mark the onset of hair formation in the composed 
Einstein-Yang-Mills-Higgs field theory. 

The Einstein-Yang-Mills-Higgs theory is characterized by the composed action 
\cite{EYMH2,VolGal}
\begin{equation}\label{Eq2}
S=d^4x\sqrt{-g}\Big[{{1}\over{16\pi G}}R-{{1}\over{16\pi}}F^2-{1\over2}(D\phi)^2-\lambda(\phi^2-v^2)^2\Big]\  .
\end{equation}
Here $R$ is the Ricci curvature scalar, 
\begin{equation}\label{Eq3}
F=dA+e[A,A]\  ,
\end{equation}
and
\begin{equation}\label{Eq4}
D\phi=d\phi+e[A,\phi]\  ,
\end{equation}
where the the Higgs field $\phi$ and the Yang-Mills connection $A$ take 
values in the Lie algebra of the $SU(2)$ gauge group \cite{EYMH2,VolGal}. 
The physical parameters $\{G,e,\lambda,v\}$ in the Einstein-Yang-Mills-Higgs action (\ref{Eq2}) 
are respectively the Newton gravity constant, the gauge coupling constant, the
Higgs coupling constant, and the vacuum expectation value of the Higgs field. 
The field theory (\ref{Eq2}) is characterized by the following two dimensionless 
physical parameters \cite{EYMH2,VolGal}: 
\begin{equation}\label{Eq5}
\alpha\equiv\sqrt{G}v\
\end{equation}
and
\begin{equation}\label{Eq6}
\beta\equiv{{\sqrt{\lambda}}\over{e}}\  .
\end{equation}

The Reissner-Nordstr\"om black-hole spacetime of mass $M$ and unit magnetic charge, which is a valid 
solution of the Einstein-Yang-Mills-Higgs equations, is characterized by the curved 
line element \cite{EYMH1,EYMH2,EYMH3,VolGal}
\begin{equation}\label{Eq7}
ds^2=-N(r)dt^2+{1\over{N(r)}}dr^2+r^2(d\theta^2+\sin^2\theta d\varphi^2)\  ,
\end{equation}
where
\begin{equation}\label{Eq8}
N(r)=1-{{2M}\over{r}}+{{1}\over{r^2}}\  .
\end{equation} 
The horizon radii of the magnetically charged Reissner-Nordstr\"om black hole, 
\begin{equation}\label{Eq9}
r_{\pm}=M\pm(M^2-1)^{1/2}\,
\end{equation}
are determined by the roots of the metric function (\ref{Eq8}). 

In order to facilitate of fully analytical treatment of the composed black-hole-field system we shall consider, 
following \cite{EYMH1,EYMH2,EYMH3,VolGal}, the case in which the 
Higgs field is frozen in its vacuum expectation value. In this case 
the spatial functional behavior of a linearized static Yang-Mills field in the curved 
black-hole spacetime (\ref{Eq7}) is determined by the 
ordinary differential equation \cite{EYMH1,EYMH2,EYMH3,VolGal}
\begin{equation}\label{Eq10}
{{d}\over{dr}}\Big[N(r){{d\psi}\over{dr}}\Big]+\Big({{1}\over{r^2}}-m_{\text{w}}^2\Big)\psi=0\  ,
\end{equation}
where the mass parameter 
\begin{equation}\label{Eq11}
m_{\text{w}}\equiv ev\
\end{equation}
of the Yang-Mills field (due to the Higgs effect) has the dimensions of length$^{-1}$. 
The spatially regular eigenfunctions of the supported Yang-Mills configurations are characterized by the physically motivated boundary conditions \cite{EYMH1,EYMH2,EYMH3,VolGal}
\begin{equation}\label{Eq12}
\psi(r=r_+)<\infty\
\end{equation}
and 
\begin{equation}\label{Eq13}
\psi(r\to\infty)\sim e^{-m_{\text{w}} r}\  .
\end{equation}

The set of equations (\ref{Eq10}), (\ref{Eq12}), and (\ref{Eq13}) 
determines the discrete resonance spectrum $\{[m_{\text{w}} r_+]_k\}_{k=0}^{k=\infty}$ of the dimensionless 
mass-radius parameter $m_{\text{w}} r_+$ that characterizes the cloudy black-hole-field configurations of the 
composed Einstein-Yang-Mills-Higgs field theory (\ref{Eq2}). 

In the next section we shall explicitly prove that the characteristic
resonance spectrum of the composed black-hole-field cloudy configurations 
can be studied analytically in the regime $1\ll r_+\ll m_{\text{w}}^{-1}$. 
In particular, we shall derive a remarkably compact analytical formula [see
Eq. (\ref{Eq35}) below] for the excited resonance spectrum $\{[m_{\text{w}} r_+]_k\}_{k=1}^{k=\infty}$ of 
the dimensionless mass-radius parameter $m_{\text{w}} r_+$ that 
characterizes the composed black-hole-field cloudy spacetimes of the Einstein-Yang-Mills-Higgs theory. 

\section{The resonance spectrum of the composed black-hole-linearized-field cloudy configurations 
of the Einstein-Yang-Mills-Higgs field theory}

In the present section we shall analyze the discrete resonance spectrum $\{[m_{\text{w}} r_+]_k\}^{k=\infty}_{k=1}$ that 
characterizes the cloudy black-hole-linearized-field bound-state configurations that mark the onset of hair 
formation in the composed Einstein-Yang-Mills-Higgs field theory (\ref{Eq2}).

Interestingly, we shall explicitly prove below that the radial differential equation (\ref{Eq10}) of the cloudy Yang-Mills configurations is amenable to an {\it analytical} treatment in the 
regime (\ref{Eq1}) of small-mass weakly-magnetized central supporting black holes. 
In particular, we shall show that the differential equation (\ref{Eq10}) can be solved analytically 
in the radial regions $r\ll m^{-1}_{\text{w}}$ and $r\gg r_+$. 
We shall then use a matching procedure in the overlapping radial region
$r_+\ll r\ll m^{-1}_{\text{w}}$ [this matching procedure would be valid for composed 
black-hole-field systems in the dimensionless regime $m_{\text{w}} r_+\ll1$, see Eq. (\ref{Eq1})] 
in order to determine the discrete resonance spectrum of the dimensionless mass-radius parameter 
\begin{equation}\label{Eq14}
\mu\equiv m_{\text{w}} r_+\
\end{equation}
that characterizes the cloudy black-hole-field configurations of the the 
composed Einstein-Yang-Mills-Higgs theory (\ref{Eq2}). 

Taking cognizance of Eqs. (\ref{Eq8}) and (\ref{Eq9}) one finds the relations 
\begin{equation}\label{Eq15}
N(r)=\Big(1-{{2M}\over{r}}\Big)\cdot\{1+O(M^{-2})\}\
\end{equation}
and 
\begin{equation}\label{Eq16}
r_+=2M\cdot\{1+O(M^{-2})\}\
\end{equation}
in the regime $M\gg1$ [see Eqs. (\ref{Eq1}) and (\ref{Eq9})] of weakly magnetized black holes. 

We shall first determine the radial functional behavior of the supported Yang-Mills configurations in the radial region
\begin{equation}\label{Eq17}
r\ll m^{-1}_{\text{w}}\  ,
\end{equation}
in which case the differential equation (\ref{Eq10}) of the linearized field can be approximated by 
\begin{equation}\label{Eq18}
{{d}\over{dr}}\Big[N(r){{d\psi}\over{dr}}\Big]+{{1}\over{r^2}}\psi=0\  .
\end{equation}
Defining the dimensionless radial coordinate 
\begin{equation}\label{Eq19}
x\equiv {{r-r_+}\over{r_+}}\  ,
\end{equation}
one finds that the physically acceptable solution of the differential equation (\ref{Eq18}) [that is, the solution 
that respects the boundary condition (\ref{Eq12}) at the horizon of the 
supporting black hole] can be expressed in terms of the hypergeometric function 
\cite{Abram,Morse,Notenm,Notehh}:
\begin{equation}\label{Eq20}
\psi(x)={_2F_1}\Big({{-1-i\sqrt{3}}\over{2}},{{-1+i\sqrt{3}}\over{2}};1;-x\Big)\  .
\end{equation}

Using Eq. 15.3.7 of \cite{Abram} one can write the radial Yang-Mills eigenfunction (\ref{Eq20}) in the form 
\begin{eqnarray}\label{Eq21}
\psi(x)={{\Gamma(i\sqrt{3})}\over{\Gamma({{-1+i\sqrt{3}}\over{2}})\Gamma({{3+i\sqrt{3}}\over{2}})}}x^{{{1+i\sqrt{3}}\over{2}}}\cdot
{_2F_1}\Big({{-1-i\sqrt{3}}\over{2}},{{-1-i\sqrt{3}}\over{2}};1-i\sqrt{3};-{{1}\over{x}}\Big)+\nonumber \\
{{\Gamma(-i\sqrt{3})}\over{\Gamma({{-1-i\sqrt{3}}\over{2}})\Gamma({{3-i\sqrt{3}}\over{2}})}}x^{{{1-i\sqrt{3}}\over{2}}}\cdot
{_2F_1}\Big({{-1+i\sqrt{3}}\over{2}},{{-1+i\sqrt{3}}\over{2}};1+i\sqrt{3};-{{1}\over{x}}\Big)\
\end{eqnarray}
which, using the simple asymptotic behavior (see Eq. 15.1.1 of \cite{Abram})
\begin{equation}\label{Eq22}
_2F_1(a,b;c;z)\to 1\ \ \ \text{for}\ \ \ {{ab}\over{c}}\cdot z\to 0\
\end{equation}
of the hypergeometric function, yields the radial functional behavior \cite{Notexr}
\begin{eqnarray}\label{Eq23}
\psi(r)={{\Gamma(i\sqrt{3})}\over{\Gamma({{-1+i\sqrt{3}}\over{2}})\Gamma({{3+i\sqrt{3}}\over{2}})}}
\cdot\Big({{r}\over{r_+}}\Big)^{{1+i\sqrt{3}}\over{2}}+
{{\Gamma(-i\sqrt{3})}\over{\Gamma({{-1-i\sqrt{3}}\over{2}})\Gamma({{3-i\sqrt{3}}\over{2}})}}\cdot \Big({{r}\over{r_+}}\Big)^{{{1-i\sqrt{3}}\over{2}}}\
\end{eqnarray}
for the characteristic eigenfunction of the cloudy Yang-Mills configurations in the intermediate region
\begin{equation}\label{Eq24}
r_+\ll r\ll m^{-1}_{\text{w}}\  .
\end{equation}

We shall now determine the radial functional behavior of the supported Yang-Mills configurations in the region
\begin{equation}\label{Eq25}
r\gg r_+\  ,
\end{equation}
in which case the differential equation (\ref{Eq10}) can be approximated by 
\begin{equation}\label{Eq26}
{{d^2\psi}\over{dr^2}}+\Big({{1}\over{r^2}}-m_{\text{w}}^2\Big)\psi=0\  .
\end{equation}
The physically acceptable solution of the differential equation (\ref{Eq26}) [that is, the field solution 
that respects the asymptotic boundary condition (\ref{Eq13}) which characterizes finite (bounded) Yang-Mills 
eigenfunctions] can be expressed in terms of 
the confluent hypergeometric function of the second kind \cite{Abram,Morse,Notejj}:
\begin{equation}\label{Eq27}
\psi(r)=A\times e^{-m_{\text{w}} r}\cdot(2m_{\text{w}} r)^{{{1+i\sqrt{3}}\over{2}}}\cdot 
U\Big({{1+i\sqrt{3}}\over{2}},1+i\sqrt{3},2m_{\text{w}} r\Big)\  ,
\end{equation}
where $A$ is a constant pre-factor. 

Using Eqs. 13.1.2 and 13.1.3 of \cite{Abram} one obtains from (\ref{Eq27}) the functional expression
\begin{eqnarray}\label{Eq28}
\psi(r)=A\times{{\pi}\over{\sin[\pi(1+i\sqrt{3})]}}\Bigg[{{(2m_{\text{w}} r)^{{{1+i\sqrt{3}}\over{2}}}}
\over{\Gamma({{1-i\sqrt{3}}\over{2}})\Gamma(1+i\sqrt{3})}}-
{{(2m_{\text{w}} r)^{{{1-i\sqrt{3}}\over{2}}}}\over{\Gamma({{1+i\sqrt{3}}\over{2}})\Gamma(1-i\sqrt{3})}}\Bigg]\
\end{eqnarray}
for the radial Yang-Mills eigenfunction in the intermediate region
\begin{equation}\label{Eq29}
r_+\ll r\ll m^{-1}_{\text{w}}\  .
\end{equation}

For composed black-hole-field configurations in the small mass-radius regime $\mu\ll1$ ($m_{\text{w}} r_+\ll1$) 
there is an overlap radial region,
\begin{equation}\label{Eq30}
r_+\ll r_{\text{o}}\ll m^{-1}_{\text{w}}\  ,
\end{equation}
where both functional expressions (\ref{Eq23}) and (\ref{Eq28}) for the radial profile 
of the linearized Yang-Mills field are valid. 
Note, in particular, that the analytically derived expressions (\ref{Eq23}) and (\ref{Eq28}), which 
determine the radial behavior of the supported Yang-Mills cloudy configurations in the overlap region (\ref{Eq30}), 
share the same functional behavior. 

Matching the radially-dependent Yang-Mills expressions
(\ref{Eq23}) and (\ref{Eq28}) in their overlap region
(\ref{Eq30}) one obtains the two dimensionless relations
\begin{eqnarray}\label{Eq31}
{{\Gamma(i\sqrt{3})}\over{\Gamma({{-1+i\sqrt{3}}\over{2}})\Gamma({{1+i\sqrt{3}}\over{2}})}}=
A\times{{\pi}\over{\sin[\pi(1+i\sqrt{3})]}}\cdot{{(2m_{\text{w}} r_+)^{{{1+i\sqrt{3}}\over{2}}}}
\over{\Gamma({{1-i\sqrt{3}}\over{2}})\Gamma(1+i\sqrt{3})}}\
\end{eqnarray}
and
\begin{eqnarray}\label{Eq32}
{{\Gamma(-i\sqrt{3})}\over{\Gamma({{-1-i\sqrt{3}}\over{2}})\Gamma({{3-i\sqrt{3}}\over{2}})}}=
-A\times{{\pi}\over{\sin[\pi(1+i\sqrt{3})]}}\cdot{{(2m_{\text{w}} r_+)^{{{1-i\sqrt{3}}
\over{2}}}}\over{\Gamma({{1+i\sqrt{3}}\over{2}})\Gamma(1-i\sqrt{3})}}\  ,
\end{eqnarray}
which yield the resonance condition
\begin{equation}\label{Eq33}
(2m_{\text{w}} r_+)^{i\sqrt{3}}=-
{{\Gamma(i\sqrt{3})\Gamma({{-1-i\sqrt{3}}\over{2}})\Gamma({{3-i\sqrt{3}}\over{2}})
\Gamma({{1-i\sqrt{3}}\over{2}})\Gamma(1+i\sqrt{3})}\over
{\Gamma(-i\sqrt{3})\Gamma({{-1+i\sqrt{3}}\over{2}})\Gamma({{1+i\sqrt{3}}\over{2}})\Gamma({{1+i\sqrt{3}}\over{2}})\Gamma(1-i\sqrt{3})}}\
\end{equation}
for the composed black-hole-field system. 

Using Eqs. 6.1.15 and 6.1.18 of \cite{Abram} one can write 
the resonance condition (\ref{Eq33}) in the compact mathematical form 
\begin{equation}\label{Eq34}
(m_{\text{w}} r_+)^{i\sqrt{3}}=2^{3i\sqrt{3}}\cdot{{1-i\sqrt{3}}\over{1+i\sqrt{3}}}\cdot
{{\Gamma(-1/2-i\sqrt{3}/2)[\Gamma(i\sqrt{3}/2)]^2}\over
{\Gamma(-1/2+i\sqrt{3}/2)[\Gamma(-i\sqrt{3}/2)]^2}}\  ,
\end{equation}
which yields the dimensionless expression \cite{Notenn} 
\begin{equation}\label{Eq35}
\mu_k=8\cdot\Bigg[{{1-i\sqrt{3}}\over{1+i\sqrt{3}}}\cdot
{{\Gamma(-1/2-i\sqrt{3}/2)[\Gamma(i\sqrt{3}/2)]^2}\over
{\Gamma(-1/2+i\sqrt{3}/2)[\Gamma(-i\sqrt{3}/2)]^2}}\Bigg]^{-{{i}\over{\sqrt{3}}}}
\times e^{-{{2\pi k}\over{\sqrt{3}}}}\ \ \ \ ; \ \ \ \ k=1,2,3,...\
\end{equation}
for the small ($\mu_k\ll1$) mass-radius resonance spectrum that characterizes 
the cloudy black-hole-field bound-state configurations of 
the composed Einstein-Yang-Mills-Higgs theory (\ref{Eq2}) \cite{Notecon}. 

\section{Summary and Discussion}

Numerical studies (see \cite{EYMH1,EYMH2,EYMH3,VolGal} and references therein) 
of the composed Einstein-Yang-Mills-Higgs field theory have established 
the physically interesting fact that spatially regular hairy field configurations can be supported by 
magnetically charged black holes. 
In particular, the critical boundary between bald black-hole spacetimes 
and composed black-hole-field hairy configurations in the Einstein-Yang-Mills-Higgs theory (\ref{Eq2}) 
is marked by the existence of cloudy spacetimes \cite{Notelc,Hod1,Hod2,Her1,Her2,Gar1} 
which describe magnetically charged Reissner-Nordstr\"om black holes 
that support linearized bound-state Yang-Mills fields. 

In the present paper we have studied the physical and mathematical properties of the
cloudy black-hole-field bound-state configurations that characterize 
the Einstein-Yang-Mills-Higgs field theory. 
Interestingly, we have explicitly proved that this composed physical system is amenable to an {\it analytical} 
treatment in the regime $1\ll r_+\ll m_{\text{w}}^{-1}$ [see Eq. (\ref{Eq1})] 
of small-mass weakly-magnetized black holes. 

The main analytical results derived in this paper and their physical implications are as follows:

(1) We have used a small mass-radius ($m_{\text{w}}r_+\ll1$) analytical matching procedure
in order to solve the radial differential equation (\ref{Eq10}) which, together with the physically 
motivated boundary conditions (\ref{Eq12}) and (\ref{Eq13}), determines the spatial 
functional behavior of the linearized Yang-Mills cloudy configurations in the magnetically charged 
Reissner-Nordstr\"om black-hole spacetime (\ref{Eq7}). 

(2) We have derived the remarkably compact expression [see Eq. (\ref{Eq35})] \cite{Note6123,Notep3}
\begin{equation}\label{Eq36}
\mu_k=8e^{{{2\phi_1+4\phi_2}\over{\sqrt{3}}}}\times e^{-{{2\pi (k+1/3)}\over{\sqrt{3}}}}\ \ \ \ 
\text{for}\ \ \ \ \mu_k\ll1
\end{equation}
for the excited resonance spectrum of the dimensionless mass-radius parameter $\mu\equiv m_{\text{w}}r_+$ 
that characterizes the cloudy black-hole-field configurations of the composed 
Einstein-Yang-Mills-Higgs field theory (\ref{Eq2}), where
\begin{equation}\label{Eq37}
\phi_1\equiv \arg[\Gamma(-1/2-i\sqrt{3}/2)]\ \ \ \ ; \ \ \ \ 
\phi_2\equiv \arg[\Gamma(i\sqrt{3}/2)]\  .
\end{equation}

(3) It is interesting to note that the analytically derived resonance formula (\ref{Eq36}) implies that the 
excited mass-radius eigenvalues of the cloudy black-hole-field bound-state configurations are characterized 
by the asymptotically constant ($k$-independent) dimensionless ratio
\begin{equation}\label{Eq38}
{{\mu_{k+1}}\over{\mu_{k}}}=e^{-{{2\pi}\over{\sqrt{3}}}}\  .
\end{equation}

(4) It is physically important to test the accuracy of the {\it analytically} derived excited resonance spectrum (\ref{Eq36}) that 
characterizes the cloudy black-hole-field configurations of the composed 
Einstein-Yang-Mills-Higgs theory (\ref{Eq2}) in the regime (\ref{Eq1}). 

To this end, we note that the exact ({\it numerically} computed \cite{EYMH3}) first excited eigenvalue 
of the dimensionless mass-radius parameter (\ref{Eq14}) is (see Fig. 6 of \cite{EYMH3})
\begin{equation}\label{Eq39}
\mu^{\text{numerical}}_{1}\simeq0.0267\  .
\end{equation}
The corresponding {\it analytically} calculated eigenvalue of the system, as obtained directly 
from the analytically derived resonance formula (\ref{Eq36}), is given by the relation 
\begin{equation}\label{Eq40}
\mu^{\text{analytical}}_{1}=8e^{{{2\phi_1+4\phi_2}\over{\sqrt{3}}}}\times e^{-{{8\pi}\over{3\sqrt{3}}}}\simeq0.0275\  .
\end{equation}
One therefore finds a fairly good agreement (to within $\sim3\%$),
\begin{equation}\label{Eq41}
{{\mu^{\text{analytical}}_{1}}\over{\mu^{\text{numrical}}_{1}}}\simeq1.03\  ,
\end{equation}
between the exact (numerically computed) first excited eigenvalue (\ref{Eq39}) of \cite{EYMH3} 
and the corresponding analytically calculated eigenvalue (\ref{Eq40}) of the mass-radius 
parameter $\mu\equiv m_{\text{w}}r_+$ that characterizes the cloudy black-hole-field bound-state 
configurations of the composed Einstein-Yang-Mills-Higgs theory.

The agreement between the numerical results and the resonance formula (\ref{Eq36}) 
is even better for higher resonances of the systems, which lie deep within the dimensionless 
regime $\mu\ll1$ [see Eqs. (\ref{Eq1}) and (\ref{Eq14})]. 
For example, the {\it numerically} computed second excited eigenvalue 
of the dimensionless mass-radius parameter is
\begin{equation}\label{Eq42}
\mu^{\text{numerical}}_{2}\simeq7.306\times10^{-4}\  .
\end{equation}
The corresponding {\it analytically} calculated second excited eigenvalue of the system, as obtained directly 
from the analytically derived resonance formula (\ref{Eq36}), is given by the relation 
\begin{equation}\label{Eq43}
\mu^{\text{analytical}}_{2}=8e^{{{2\phi_1+4\phi_2}\over{\sqrt{3}}}}\times e^{-{{14\pi}\over{3\sqrt{3}}}}
\simeq7.309\times10^{-4}\  .
\end{equation}
One therefore finds a remarkably good agreement (to within $\sim0.04\%$),
\begin{equation}\label{Eq44}
{{\mu^{\text{analytical}}_{2}}\over{\mu^{\text{numrical}}_{2}}}\simeq1.0004\  ,
\end{equation}
between the exact (numerically computed) second excited eigenvalue (\ref{Eq42}) 
and the corresponding analytically calculated second excited eigenvalue (\ref{Eq43}) of the composed 
black-hole-field system.

\bigskip
\noindent
{\bf ACKNOWLEDGMENTS}
\bigskip

This research is supported by the Carmel Science Foundation. I would
like to thank Yael Oren, Arbel M. Ongo, Ayelet B. Lata, and Alona B.
Tea for helpful discussions.



\begin{thebibliography}{99}
\bibitem{Notelc} The expression ‘linearized clouds' is commonly used in the physics literature (see \cite{Hod1,Hod2,Her1,Her2,Gar1} and references therein) to describe linearized bound-state fields which are 
supported in curved black-hole spacetimes.
\bibitem{Notema} Note that the mass parameter $m_{\text{w}}$ which is used in the present paper 
corresponds to the parameter $\alpha\equiv m_{\text{w}}/gm_{\text{p}}$ which is used in \cite{EYMH1,EYMH2}, 
where $g$ is the gauge coupling of the theory and $m_{\text{p}}$ is the Planck mass.
\bibitem{Noteunits} We use natural units in which $G=c=e=1$.
\bibitem{NoteP1} We shall consider weakly-magnetized central supporting Reissner-Nordstr\"om
black holes with unit magnetic charge which is small on the
scale set by the black-hole radius $r_+$.
\bibitem{Noteannu} Note that the assumed strong inequality (\ref{Eq1}) is physically motivated by the numerical results presented in \cite{EYMH3} which indicate that the excited eigenvalues of the system 
lie deep within the dimensionless regime (\ref{Eq1}).
\bibitem{Notewa1} It is worth noting that while numerical results provide accurate data, 
an approximate analytical formula is highly valuable because it highlights the dependence of the characteristic 
eigenvalues of the system on the discrete resonance parameter $k$ [see Eq. (\ref{Eq36}) below]. 
In particular, the analytically derived resonance formula (\ref{Eq36}) reveals the interesting fact that 
the excited eigenvalues of the composed black-hole-field system (\ref{Eq2}) 
converge exponentially fast to zero.
\bibitem{Notewa2} It is worth emphasizing that, 
due to the rapid (exponential) convergence of the resonant eigenvalues to zero, the growing and 
decaying modes of the numerical solution (which asymptotically behave as $e^{m_{\text{w}}r}$ 
and $e^{-m_{\text{w}}r}$, respectively) separate only very slowly. 
Consequently, in order to make $e^{m_{\text{w}}r}$ and $e^{-m_{\text{w}}r}$ numerically distinguishable, 
one must employ extremely large computational grids extending to $r_{\text{max}} \gg m_{\text{w}}^{-1} \gg r_{+}$. 
For example, an accurate numerical computation of the second ($k=2$) excited eigenvalue of the system requires an integration 
grid with $r_{\text{max}}\gtrsim 10^4r_{+}$, whereas the computation of the third ($k=3$) 
excited eigenvalue of the system already demands an integration grid with $r_{\text{max}}\gtrsim 10^6r_{+}$. 
Thus, the numerical integration of the ordinary differential equation (\ref{Eq10}) 
becomes exponentially more expensive as the resonance parameter $k$ increases. 
This is precisely where the analytically derived resonance formula (\ref{Eq36}) proves both most necessary and most accurate.
\bibitem{Notenm} The normalization factor of the analytically derived 
linearized Yang-Mills function (\ref{Eq20}) is chosen, without loss of generality, to be $1$.
\bibitem{Notehh} Using the characteristic asymptotic behavior $_2F_1(a,b;c;z)\to 1$ for $(ab/c)\cdot z\to 0$
of the hypergeometric function (see Eq. 15.1.1 of \cite{Abram}) one deduces 
that the Yang-Mills radial eigenfunction (\ref{Eq20}) is characterized by 
the near-horizon behavior $\psi(x\to0)\to1$.
\bibitem{Notexr} Here we have used the leading-order dimensionless relation $x=r/r_++O(1)$ for 
$r\gg r_+$ [see Eq. (\ref{Eq19})].
\bibitem{Notejj} Using the characteristic asymptotic behavior $U(a,b,z)\to z^{-a}[1+O(z^{-1})]$ 
for $z\to\infty$ of the confluent hypergeometric function (see Eq. 13.1.7 of \cite{Abram}) 
one deduces that the Yang-Mills radial eigenfunction (\ref{Eq27}) is characterized by the 
asymptotic large-$r$ functional behavior $\psi(r\to\infty)\sim e^{-m_{\text{w}} r}$.
\bibitem{Notenn} Here we have used the relation $1=e^{-i2\pi k}$, 
where the integer $k=1,2,3,...$ is the dimensionless resonance parameter that characterizes the 
cloudy black-hole-field bound-state configurations of the Einstein-Yang-Mills-Higgs theory.
\bibitem{Notecon} It is worth noting that, using Eq. 6.1.23 of \cite{Abram}, 
one finds the relation ${{1-i\sqrt{3}}\over{1+i\sqrt{3}}}\cdot
{{\Gamma(-1/2-i\sqrt{3}/2)[\Gamma(i\sqrt{3}/2)]^2}\over
{\Gamma(-1/2+i\sqrt{3}/2)[\Gamma(-i\sqrt{3}/2)]^2}}=e^{i\theta}$ [see Eq. (\ref{Eq35})] 
with $\theta\in\mathbb{R}$, which implies that $\mu_k\in\mathbb{R}$.
\bibitem{Notep3} Here we have used the relation $\arg(1-i\sqrt{3})=-\pi/3$.
\end{thebibliography}
\end{document}